\documentclass[dvips,12pt]{article}

\usepackage{cite,amsmath,epsfig}   
\usepackage{amssymb}
\graphicspath{{smallfigs/}}

\def\beq{\begin{equation}}
\def\eeq{\end{equation}}

\def\bea{\begin{eqnarray}}
\def\eea{\end{eqnarray}}
\def\ba{\begin{array}}             
\def\ea{\end{array}}

\newcommand{\FUTA}{{\bf FUTA}}
\newcommand{\FUTB}{{\bf FUTB}}

\def\refeq#1{\mbox{eq.~(\ref{#1})}}
\def\refeqs#1{\mbox{eqs.~(\ref{#1})}}
\def\reffi#1{\mbox{Fig.~\ref{#1}}}
\def\reffis#1{\mbox{Figs.~\ref{#1}}}
\def\refta#1{\mbox{Tab.~\ref{#1}}}
\def\refse#1{\mbox{Sect.~\ref{#1}}}
\def\refses#1{\mbox{Sects.~\ref{#1}}}
\def\citere#1{\mbox{Ref.~\cite{#1}}}
\def\citeres#1{\mbox{Refs.~\cite{#1}}}

\newcommand{\mste}{m_{\tilde{t}_1}}
\newcommand{\mstz}{m_{\tilde{t}_2}}

\newcommand{\msbe}{m_{\tilde{b}_1}}
\newcommand{\msbz}{m_{\tilde{b}_2}}

\newcommand{\At}{A_t}

\newcommand{\msusy}{M_{\rm SUSY}}

\newcommand{\SU}{\mathrm {SUSY}}

\def\order#1{${\cal O}(#1)$}

\newcommand{\cp}{{\cal CP}}

\def\ed#1{\frac{1}{#1}}

\newcommand{\MZ}{M_Z}
\newcommand{\MA}{M_A}

\newcommand{\Mh}{M_h}
\newcommand{\MH}{M_H}

\newcommand{\MHp}{M_{H^\pm}}

\newcommand{\mt}{m_{t}}
\newcommand{\Mt}{m_{t}}

\newcommand{\mb}{m_{b}}
\newcommand{\mbbar}{\overline{m}_b}

\newcommand{\mgl}{m_{\tilde{g}}}

\newcommand{\Sbot}{\tilde{b}}

\newcommand{\Smu}{\tilde \mu}
\newcommand{\Sneum}{\tilde \nu_\mu}

\newcommand{\tsf}{\theta\kern-.20em_{\tilde{f}}}
\newcommand{\tsfp}{\theta\kern-.20em_{\tilde{f}\prime}}
\newcommand{\tsq}{\theta\kern-.15em_{\tilde{q}}}

\newcommand{\cha}[1]{\tilde \chi^\pm_{#1}}
\newcommand{\mcha}[1]{m_{\tilde \chi^\pm_{#1}}}
\newcommand{\neu}[1]{\tilde \chi^0_{#1}}
\newcommand{\mneu}[1]{m_{\tilde \chi^0_{#1}}}

\newcommand{\KL}{\left(}
\newcommand{\KR}{\right)}

\newcommand{\VL}{\left( \begin{array}{c}}
\newcommand{\VR}{\end{array} \right)}
\newcommand{\ML}{\left( \begin{array}{cc}}
\newcommand{\MLd}{\left( \begin{array}{ccc}}
\newcommand{\MLv}{\left( \begin{array}{cccc}}
\newcommand{\MR}{\end{array} \right)}

\newcommand{\Tb}{\tan \beta\hspace{1mm}}
\newcommand{\tb}{\tan \beta}

\newcommand{\tev}{\,\, \mathrm{TeV}}
\newcommand{\gev}{\,\, \mathrm{GeV}}

\newcommand{\BC}{\begin{center}}
\newcommand{\EC}{\end{center}}
\newcommand{\BE}{\begin{equation}}
\newcommand{\EE}{\end{equation}}
\newcommand{\BEA}{\begin{eqnarray}}
\newcommand{\BEAnn}{\begin{eqnarray*}}
\newcommand{\EEA}{\end{eqnarray}}
\newcommand{\EEAnn}{\end{eqnarray*}}
\newcommand{\non}{\nonumber}
\newcommand{\id}{{\rm 1\kern-.12em
\rule{0.3pt}{1.5ex}\raisebox{0.0ex}{\rule{0.1em}{0.3pt}}}}
\newcommand{\lsim}
{\;\raisebox{-.3em}{$\stackrel{\displaystyle <}{\sim}$}\;}
\newcommand{\gsim}
{\;\raisebox{-.3em}{$\stackrel{\displaystyle >}{\sim}$}\;}

\def\al{\alpha}

\def\als{\alpha_s}
\def\alt{\alpha_t}

\def\ga{\gamma}

\def\si{\sigma}

\def\De{\Delta}

\newcommand{\ifb}{$\mbox{fb}^{-1}$}

\newcommand{\br}{{\rm BR}}

\newcommand{\amu}{a_\mu}

\newcommand{\amuexp}{a_\mu^{\rm exp}}
\newcommand{\amutheo}{a_\mu^{\rm theo}}

\evensidemargin \oddsidemargin
\begin{document}

\thispagestyle{empty}

\def\thefootnote{\fnsymbol{footnote}}

\begin{flushright}
arXiv:0712.3630 [hep-ph]
\end{flushright}

\vspace{1cm}

\begin{center}
{\large\sc {\bf Confronting Finite Unified Theories\\[.5em] 
with Low-Energy Phenomenology}}

\vspace{0.4cm}

\vspace{1cm}

{\sc 
S.~Heinemeyer$^{1}$%
\footnote{email: Sven.Heinemeyer@cern.ch}%
, M~Mondrag\'on$^{2}$%
\footnote{email: myriam@fisica.unam.mx}%
~and G.~Zoupanos$^{3}$%
\footnote{email: George.Zoupanos@cern.ch}
}

\vspace*{1cm}

{\sl

$^1$Instituto de FM-msica de Cantabria (CSIC-UC),\\
Edificio Juan Jorda,
Avda. de Los Castros s/n\\
39005 Santander, Spain

\vspace*{0.4cm}

$^2$Inst.~de FM-msica, Universidad~Nacional Aut\'onoma de M\'exico,\\
Apdo. Postal 20-364,   M\'exico 01000 D.F., M\'exico

\vspace*{0.4cm}

$^3$Physics Department, National Technical University of Athens,\\
Zografou Campus: Heroon Polytechniou 9,\\
15780 Zografou, Athens, Greece

}
\end{center}

\vspace*{0.2cm}
\begin{abstract}
  Finite Unified Theories (FUTs) are $N=1$ supersymmetric Grand Unified
  Theories that can be made all-loop finite. The requirement of
  all-loop finiteness leads to a severe reduction of the free
  parameters of the theory and, in turn, to a large number of
  predictions.  FUTs are investigated in the context of low-energy
  phenomenology observables. We present a detailed
  scanning of the all-loop finite 
  $SU(5)$~FUTs, where we include the  theoretical uncertainties at
  the unification scale and we apply several phenomenological constraints.
  Taking into account the restrictions
  from the top and bottom quark masses, we can discriminate between
  different models. Including further low-energy constraints such as
  $B$~physics observables, the bound on the lightest Higgs boson mass
  and the cold dark matter density, we determine the predictions of
  the allowed parameter space for the Higgs boson sector and the
  supersymmetric particle spectrum of the selected model.
\end{abstract}

\def\thefootnote{\arabic{footnote}}
\setcounter{page}{0}
\setcounter{footnote}{0}

\newpage


\section{Introduction}

A large and sustained effort has been done in the recent years aiming
to achieve a unified description of all interactions. Out of this
endeavor two main directions have emerged as the most promising to
attack the problem, namely, the superstring theories and 
non-commutative geometry. The two approaches, although at a different
stage of development, have common unification targets and share similar
hopes for exhibiting improved renormalization properties in the
ultraviolet(UV) as compared to ordinary field theories.  Moreover the
two frameworks came closer by the observation that a natural
realization of non-commutativity of space appears in the string theory
context of D-branes in the presence of a constant background
antisymmetric field \cite{Connes:1997cr}.  However,
despite the importance of having frameworks to discuss quantum gravity
in a self-consistent way and possibly to construct there finite
theories, it is very interesting to search for the minimal realistic
framework in which finiteness can take place.
In addition the main goal expected from a unified
description of interactions by the particle physics community is to
understand the present day large number of free parameters of the
Standard Model (SM) in terms of a few fundamental ones. In other
words, to achieve {\it reduction of couplings} at a more fundamental
level.  A complementary, and certainly not contradicting, program
has been developed \cite{finite1,kkmz1,acta} in
searching for a more fundamental theory possibly at the Planck scale
called Finite Unified Theories (FUTs), whose basic ingredients are
field theoretical Grand Unified Theories (GUTs) and supersymmetry (SUSY), but
its consequences certainly go beyond the known ones.

Finite Unified Theories are $N=1$ supersymmetric 
GUTs which can be made finite even to all-loop orders,
including the soft supersymmetry breaking sector.  The method to
construct GUTs with reduced independent parameters
\cite{zoup-kmz1,zoup-zim1} consists of searching for renormalization
group invariant (RGI) relations holding below the Planck scale, which
in turn are preserved down to the GUT scale. Of particular interest is
the possibility to find RGI relations among couplings that guarantee
finiteness to all-orders in perturbation theory
\cite{zoup-lucchesi1,zoup-ermushev1}.  In order to achieve the latter
it is enough to study the uniqueness of the solutions to the
one-loop finiteness conditions
\cite{zoup-lucchesi1,zoup-ermushev1,soft}.  The constructed 
{\it finite unified} $N=1$ supersymmetric GUTs, using the above tools,
predicted correctly from the dimensionless sector (Gauge-Yukawa
unification), among others, the top quark mass \cite{finite1}.  The
search for RGI relations and finiteness has been extended to the soft
supersymmetry breaking sector (SSB) of these theories
\cite{kmz2,kkk1,kkz,zoup-jack1,jack4,jack2,hisano1,yamada1,Jack:1994rk,kazakov1}, which involves parameters of dimension one and two. 
Eventually, the full theories can be made all-loop finite and their
predictive power is extended to the Higgs sector and the
SUSY spectrum. This, in turn, allows to make
predictions for low-energy precision and astrophysical
observables.  The purpose of the present article is to do an
exhaustive search of these latter predictions of the SU(5) finite
models, taking into account the restrictions resulting from the
low-energy observables.  Then we present the predictions of the models
under study for the parameter space that is still allowed after taking
the phenomenological restrictions into account. Here we focus on
the Higgs boson sector and the SUSY spectrum.

In our search we consider the restrictions imposed on the
parameter space of the models due to the following observables: 
the 3rd generation quark masses, rare
$b$~decays, $\br(b \to s \ga)$ and $\br(B_s \to \mu^+ \mu^-)$, as well as
the mass of the lightest $\cp$-even Higgs boson, $\Mh$. 
Present data on these observables already provide interesting
information about the allowed SUSY mass scales.  The non-discovery of
the Higgs boson at LEP~\cite{LEPHiggsSM,LEPHiggsMSSM} excludes a part
of the otherwise allowed parameter space.  However the non-discovery
of supersymmetric particles at LEP does not impose any restrictions on
the parameter space of the models, given that their SUSY spectra turn
out to be very heavy anyway.  An important further constraint is
provided by the density of dark matter in the Universe, which is
tightly constrained by WMAP and other astrophysical and cosmological
data~\cite{WMAP}, assuming that the dark matter consists largely of
neutralinos~\cite{EHNOS}. We also briefly discuss the implication from 
the anomalous magnetic moment of the muon, \mbox{$(g-2)_\mu$}.
Other recent analyses of GUT based models confronted with low-energy
observables and dark matter constraints can be found in 
\citeres{other,ehoww}.

In this context we first review the sensitivity of each observable to
indirect effects of supersymmetry, taking into account the present
experimental and theoretical uncertainties.  Later on we investigate
the part of parameter space in the FUT models under consideration that
is still allowed taking into account all low-energy observables.

In \refse{sec:finiteness} of the paper we review the conditions of
finiteness in $N=1$ SUSY gauge theories. The consequences of
finiteness for the soft SUSY-breaking terms are discussed in
\refse{sec:ssb}. The two $SU(5)$ FUT models that emerge are briefly
presented in \refse{sec:futs}. In \refse{sec:ewpo} we discuss
different precision observables, including the cold dark matter
constraint. \refse{sec:finalpred} contains the analysis of the parts
of parameter space that survive all constraints and the final
predictions of the models. We conclude with \refse{sec:conclusions}.


\section{Reduction of Couplings and Finiteness in $N=1$ 
   SUSY Gauge Theories} 
\label{sec:finiteness}

Here we review the main points and ideas concerning the 
{\it reduction of couplings} and {\it finiteness} in $N=1$ supersymmetric
theories.  A RGI relation among couplings $g_i$, 
$\Phi(g_1,\cdots,g_N) \mbox{=0}$, has to satisfy the partial differential
equation $ \mu\;d \Phi /d \mu = \sum_{i=1}^{N} \,\beta_{i}\,\partial
\Phi /\partial g_{i} \mbox{=\,0}, $ where $\beta_i$ is the $\beta$-function of
$g_i$.  There exist ($N-1$) independent $\Phi$'s, and finding the
complete set of these solutions is equivalent to solve the so-called
reduction equations (REs) \cite{zoup-kmz1}, $ \beta_{g} \,(d g_{i}/d
g) =\beta_{i}~,~i=1,\ldots,N, $ where $g$ and $\beta_{g}$ are the
primary coupling (in favor of which all other couplings are reduced)
and its $\beta$-function.  Using all the $(N-1)\,\Phi$'s to impose RGI
relations, one can in principle express all the couplings in terms of
a single coupling $g$.  The complete reduction, which formally
preserves perturbative renormalizability, can be achieved by demanding
a power series solution, whose uniqueness can be investigated at the
one-loop level.
 
Finiteness can be understood by considering a chiral, anomaly free,
$N~=~1$ globally supersymmetric gauge theory based on a group G with
gauge coupling constant $g$. The superpotential of the theory is given
by
\begin{equation}
 W= \frac{1}{2}\,m^{ij} \,\Phi_{i}\,\Phi_{j}+
\frac{1}{6}\,C^{ijk} \,\Phi_{i}\,\Phi_{j}\,\Phi_{k}~, 
\label{supot}
\end{equation}
where $m^{ij}$ (the mass terms) and $C^{ijk}$ (the Yukawa couplings)
are gauge invariant tensors and the matter field $\Phi_{i}$ transforms
according to the irreducible representation $R_{i}$ of the gauge group
$G$.  

 The one-loop $\beta$-function of the gauge
coupling $g$ is given by 
\bea
\beta^{(1)}_{g}&=&\frac{d g}{d t} =
\frac{g^3}{16\pi^2}\,[\,\sum_{i}\,\ell(R_{i})-3\,C_{2}(G)\,]~,
\label{betag}
\eea
where $\ell (R_{i})$ is the Dynkin index of $R_{i}$ and $C_{2}(G)$
 is the
quadratic Casimir of the adjoint representation of the
gauge group $G$. The $\beta$-functions of
$C^{ijk}$,
by virtue of the non-renormalization theorem, are related to the
anomalous dimension matrix $\gamma^j_i$ of the matter fields
$\Phi_{i}$ as
\beq
\beta_{C}^{ijk}=\frac{d}{dt}\,C^{ijk}
=C^{ijp}\,
\sum_{n=1}\frac{1}{(16\pi^2)^n}\,\gamma_{p}^{k(n)} +(k
\leftrightarrow i) +(k\leftrightarrow j)~.
\label{betay}
\eeq
At one-loop level $\gamma^j_i$ is given by 
\bea
\gamma_i^{j(1)}=\frac{1}{2}C_{ipq}\,C^{jpq}-2\,g^2\,C_{2}(R_{i})\delta_i^j~,
\label{gamay}
\eea
where $C_{2}(R_{i})$ is the quadratic Casimir of the representation
$R_{i}$, and $C^{ijk}=C_{ijk}^{*}$.

All the one-loop $\beta$-functions of the theory vanish if the
$\beta$-function of the gauge coupling $\beta_g^{(1)}$, and the
anomalous dimensions $\gamma_i^{j(1)}$,
vanish, i.e.
\begin{equation}
\sum _i \ell (R_i) = 3 C_2(G) \,,~
\frac{1}{2}C_{ipq} C^{jpq} = 2\delta _i^j g^2  C_2(R_i) .
\label{zoup-fini}
\end{equation}

A very interesting result is that the conditions (\ref{zoup-fini}) are
necessary and sufficient for finiteness at
the two-loop level \cite{soft,zoup-jack1}.

The one- and two-loop finiteness conditions (\ref{zoup-fini}) restrict
considerably the possible choices of the irreducible representations
$R_i$ for a given group $G$ as well as the Yukawa couplings in the
superpotential (\ref{supot}).  Note in particular that the finiteness
conditions cannot be applied to the supersymmetric standard model
(SSM). The presence of a $U(1)$ gauge group, whose $C_2[U(1)]=0$,
makes impossible to satisfy the condition (\ref{zoup-fini}).  This
leads to the expectation that finiteness should be attained at the
grand unified level only, the SSM being just the corresponding
low-energy, effective theory.

The finiteness conditions impose relations between gauge and Yukawa
couplings.  Therefore, we have to guarantee that such relations
leading to a reduction of the couplings hold at any renormalization
point.  The necessary, but also sufficient, condition for this to
happen is to require that such relations are solutions to the
reduction equations (REs) to all orders.  The all-loop order
finiteness theorem of \cite{zoup-lucchesi1} is based on: (a) the
structure of the supercurrent in $N=1$ SYM and on (b) the
non-renormalization properties of $N=1$ chiral anomalies.
Alternatively, similar results can be obtained
\cite{zoup-ermushev1,zoup-strassler} using an analysis of the all-loop
NSVZ gauge beta-function \cite{zoup-novikov1}.


\section{Soft supersymmetry breaking and finiteness}
\label{sec:ssb}

The above described method of reducing the dimensionless couplings has
been extended~\cite{kmz2} to the soft supersymmetry breaking (SSB)
dimensionful parameters of $N=1$ supersymmetric theories. In addition
it was found~\cite{kkk1} that RGI SSB scalar masses in general
Gauge-Yukawa unified models satisfy a universal sum rule at one-loop,
which was subsequently extended first up to two-loops~\cite{kkmz1} and
then to all-loops~\cite{kkz}.

To be more specific, consider the superpotential given
by~(\ref{supot}) along with the Lagrangian for SSB terms  
\begin{equation} 
-{\cal L}_{\rm SB} = \frac{1}{6} \,h^{ijk}\,\phi_i \phi_j \phi_k +
\frac{1}{2} \,b^{ij}\,\phi_i \phi_j + \frac{1}{2}
\,(m^2)^{j}_{i}\,\phi^{*\,i} \phi_j+ \frac{1}{2} \,M\,\lambda
\lambda+\mbox{h.c.}\,,
\end{equation} 
where the $\phi_i$ are the scalar parts of the chiral superfields
$\Phi_i$ , $\lambda$ are the gauginos and $M$ their unified mass.
Since we would like to consider only finite theories here, we assume
that the one-loop $\beta$-function of the gauge coupling $g$ vanishes.
We also assume that the reduction equations admit power series
solutions of the form $C^{ijk} = g\,\sum_{n=0}\,\rho^{ijk}_{(n)}
g^{2n}~.  $ According to the finiteness theorem of
ref.~\cite{zoup-lucchesi1}, the theory is then finite to all orders in
perturbation theory, if, among others, the one-loop anomalous
dimensions $\gamma_{i}^{j(1)}$ vanish.  The one- and two-loop
finiteness for $h^{ijk}$ can be achieved~\cite{soft,zoup-jack1} by
imposing the condition
\begin{equation} 
h^{ijk} = -M C^{ijk}+\cdots =-M \rho^{ijk}_{(0)}\,g+O(g^5)\,.
\label{hY}
\end{equation}
In addition, it was found~\cite{kkmz1} that one and two-loop
finiteness requires that the following two-loop sum rule for the soft
scalar masses has to be satisfied 
\begin{equation}
  \frac{(~m_{i}^{2}+m_{j}^{2}+m_{k}^{2}~)}{M M^{\dag}} =
  1+\frac{g^2}{16 \pi^2}\,\Delta^{(2)} +O(g^4)\,,
\label{zoup-sumr} 
\end{equation}
where $\Delta^{(2)}$ is the two-loop correction,
\begin{equation}
\Delta^{(2)} = -2\sum_{i} \left[\left(\frac{m^{2}_{i}}{M
    M^{\dag}}\right)-\left(\frac{1}{3}\right)\right]~\ell (R_i)\,,
\label{delta}
\end{equation}
which vanishes for the universal choice~\cite{zoup-jack1}, as well as
in the models we consider in the next section.
Furthermore, it was found~\cite{jack4} that the relation
\begin{equation}
h^{ijk} = -Mg (C^{ijk})' \equiv -Mg \frac{d C^{ijk}(g)}{d \ln g}\,,
\label{zoup-hmc}
\end{equation} 
among couplings is all-loop RGI. Moreover, the progress made using the
spurion technique leads to all-loop relations among SSB
$\beta$-functions \cite{acta,jack4}
and~\cite{hisano1,yamada1,Jack:1994rk,kazakov1}, which allowed to find the
all-loop RGI sum rule~\cite{kkz} in the
Novikov-Shifman-Vainstein-Zakharov scheme~\cite{zoup-novikov1}.


\section{Finite Unified Theories}
\label{sec:futs}

Finite Unified Theories (FUTs) have always attracted interest for their
intriguing mathematical properties and their predictive power. 
One very important result is that the one-loop finiteness conditions
(\ref{zoup-fini}) are sufficient to guarantee two-loop finiteness
\cite{Parkes:1984dh}.  A classification of possible one-loop finite
models was done independently by several authors \cite{Hamidi:1984ft}.
The first one and two-loop finite $SU(5)$ model was presented in
\cite{Jones:1984qd}, and shortly afterwards the conditions for
finiteness in the soft SUSY-breaking sector at one-loop
\cite{soft} were given.  In \cite{Leon:1985jm} a one and
two-loop finite $SU(5)$ model was presented where the rotation of the
Higgs sector was proposed as a way of making it realistic.  The first
all-loop finite theory was studied in \cite{finite1}, without
taking into account the soft breaking terms. Finite soft breaking
terms and the proof that one-loop finiteness in the soft terms also
implies two-loop finiteness was done in \cite{zoup-jack1}.  The
inclusion of soft breaking terms in a realistic model was done in
\cite{Kazakov:1995cy} and their finiteness to all-loops studied in
\cite{Kazakov:1997nf}, although the universality of the soft breaking
terms lead to a charged LSP. This fact was also noticed in
\cite{Yoshioka:1997yt}, where the inclusion of an extra parameter in
the Higgs sector was introduced to alleviate it. The derivation of the
sum-rule in the soft supersymmetry breaking sector and the proof that
it can be made all-loop finite were done in
\cite{kkz,zoup-kkmz1,Jones:1984qd,Leon:1985jm}, allowing thus for the
construction of 
all-loop finite realistic models.

From the classification of theories with vanishing one-loop gauge
$\beta$ function \cite{Hamidi:1984ft}, one can easily see that there
exist only two candidate possibilities to construct $SU(5)$ GUTs with
three generations. These possibilities require that the theory should
contain as matter fields the chiral supermultiplets ${\bf
  5},~\overline{\bf 5},~{\bf 10}, ~\overline{\bf 5},~{\bf 24}$ with
the multiplicities $(6,9,4,1,0)$ and $(4,7,3,0,1)$, respectively. Only
the second one contains a ${\bf 24}$-plet which can be used to provide
the spontaneous symmetry breaking (SB) of $SU(5)$ down to $SU(3)\times
SU(2)\times U(1)$. For the first model one has to incorporate another
way, such as the Wilson flux breaking mechanism to achieve the desired
SB of $SU(5)$ \cite{finite1}. Therefore, for a self-consistent field
theory discussion we would like to concentrate only on the second
possibility.

The particle content of the models we will study consists of the
following supermultiplets: three ($\overline{\bf 5} + \bf{10}$),
needed for each of the three generations of quarks and leptons, four
($\overline{\bf 5} + {\bf 5}$) and one ${\bf 24}$ considered as Higgs
supermultiplets. 
When the gauge group of the finite GUT is broken the theory is no
longer finite, and we will assume that we are left with the MSSM.

Therefore, a predictive Gauge-Yukawa unified $SU(5)$
model which is finite to all orders, in addition to the requirements
mentioned already, should also have the following properties:

\begin{enumerate}

\item 
One-loop anomalous dimensions are diagonal,
i.e.,  $\gamma_{i}^{(1)\,j} \propto \delta^{j}_{i} $.
\item The three fermion generations, in the irreducible representations
  $\overline{\bf 5}_{i},{\bf 10}_i~(i=1,2,3)$,  should
  not couple to the adjoint ${\bf 24}$.
\item The two Higgs doublets of the MSSM should mostly be made out of a
pair of Higgs quintet and anti-quintet, which couple to the third
generation.
\end{enumerate}

In the following we discuss two versions of the all-order finite
model.  The model of \citere{finite1}, which will be labeled ${\bf
  A}$, and a slight variation of this model (labeled ${\bf B}$), which
can also be obtained from the class of the models suggested in 
\citere{zoup-avdeev1} with a modification to suppress non-diagonal
anomalous dimensions \cite{kkmz1}.

The superpotential which describes the two models before the reduction
of couplings takes places is of the form
\cite{finite1,zoup-kkmz1,Jones:1984qd,Leon:1985jm} 
\bea 
W &=&
\sum_{i=1}^{3}\,[~\frac{1}{2}g_{i}^{u} \,{\bf 10}_i{\bf 10}_i H_{i}+
g_{i}^{d}\,{\bf 10}_i \overline{\bf 5}_{i}\, \overline{H}_{i}~] +
g_{23}^{u}\,{\bf 10}_2{\bf 10}_3 H_{4} \label{zoup-super1}\\
& &+g_{23}^{d}\,{\bf 10}_2 \overline{\bf 5}_{3}\, \overline{H}_{4}+
g_{32}^{d}\,{\bf 10}_3 \overline{\bf 5}_{2}\, \overline{H}_{4}+
\sum_{a=1}^{4}g_{a}^{f}\,H_{a}\, {\bf 24}\,\overline{H}_{a}+
\frac{g^{\lambda}}{3}\,({\bf 24})^3~,\nonumber
\eea
where 
$H_{a}$ and $\overline{H}_{a}~~(a=1,\dots,4)$
stand for the Higgs quintets and anti-quintets.

We will investigate two realizations of the model, labelled {\bf A} and
  {\bf B}. The main difference between model ${\bf A}$ and model ${\bf B}$ is
that two pairs of Higgs quintets and anti-quintets couple to the ${\bf
  24}$ in ${\bf B}$, so that it is not necessary to mix them with
$H_{4}$ and $\overline{H}_{4}$ in order to achieve the triplet-doublet
splitting after the symmetry breaking of $SU(5)$ \cite{kkmz1}.  Thus,
although the particle content is the same, the solutions to
Eq.(\ref{zoup-fini}) and the sum rules are different, which will
reflect in the phenomenology, as we will see.

\subsection{\FUTA }
\begin{table}
\begin{center}
\renewcommand{\arraystretch}{1.3}
\begin{tabular}{|l|l|l|l|l|l|l|l|l|l|l|l|l|l|l|l|}
\hline
& $\overline{{\bf 5}}_{1} $ & $\overline{{\bf 5}}_{2} $& $\overline{{\bf
    5}}_{3}$ & ${\bf 10}_{1} $ &  ${\bf 10}_{2}$ & ${\bf
  10}_{3} $ & $H_{1} $ & $H_{2} $ & $H_{3} $ &$H_{4 }$&  $\overline H_{1} $ &
$\overline H_{2} $ & $\overline H_{3} $ &$\overline H_{4 }$& ${\bf 24} $\\ \hline
$Z_7$ & 4 & 1 & 2 & 1 & 2 & 4 & 5 & 3 & 6 & -5 & -3 & -6 &0& 0 & 0 \\\hline
$Z_3$ & 0 & 0 & 0 & 1 & 2 & 0 & 1 & 2 & 0 & -1 & -2 & 0 & 0 & 0&0  \\\hline
$Z_2$ & 1 & 1 & 1 & 1 & 1 & 1 & 0 & 0 & 0 & 0 & 0 & 0 &  0 & 0 &0 \\\hline
\end{tabular}
  \caption{Charges of the $Z_7\times Z_3\times Z_2$ symmetry for Model
    \FUTA. }
\renewcommand{\arraystretch}{1.0}
\label{tableA}
\end{center}
\end{table}

After the reduction of couplings 
the symmetry of the superpotential $W$ (\ref{zoup-super1}) is enhanced.
For  model ${\bf A}$ one finds that
the superpotential has the
$Z_7\times Z_3\times Z_2$ discrete symmetry with the charge assignment
as shown in Table \ref{tableA}, and with the following superpotential
\bea
W &=& \sum_{i=1}^{3}\,[~\frac{1}{2}g_{i}^{u}
\,{\bf 10}_i{\bf 10}_i H_{i}+
g_{i}^{d}\,{\bf 10}_i \overline{\bf 5}_{i}\,
\overline{H}_{i}~] +
g_{4}^{f}\,H_{4}\, 
{\bf 24}\,\overline{H}_{4}+
\frac{g^{\lambda}}{3}\,({\bf 24})^3~,
\label{w-futa}
\eea

The non-degenerate and isolated solutions to $\gamma^{(1)}_{i}=0$ for
 model \FUTA, which are the boundary conditions for the Yukawa
 couplings at the GUT scale, are: 
\bea 
&& (g_{1}^{u})^2
=\frac{8}{5}~g^2~, ~(g_{1}^{d})^2
=\frac{6}{5}~g^2~,~
(g_{2}^{u})^2=(g_{3}^{u})^2=\frac{8}{5}~g^2~,\label{zoup-SOL5}\\
&& (g_{2}^{d})^2 = (g_{3}^{d})^2=\frac{6}{5}~g^2~,~
(g_{23}^{u})^2 =0~,~
(g_{23}^{d})^2=(g_{32}^{d})^2=0~,
\nonumber\\
&& (g^{\lambda})^2 =\frac{15}{7}g^2~,~ (g_{2}^{f})^2
=(g_{3}^{f})^2=0~,~ (g_{1}^{f})^2=0~,~
(g_{4}^{f})^2= g^2~.\nonumber 
\eea 
In the dimensionful sector, the sum rule gives us the following
boundary conditions at the GUT scale for this model
\cite{zoup-kkmz1,Jones:1984qd,Leon:1985jm}: 
\bea
m^{2}_{H_u}+
2  m^{2}_{{\bf 10}} &=&
m^{2}_{H_d}+ m^{2}_{\overline{{\bf 5}}}+
m^{2}_{{\bf 10}}=M^2 ~~,
\eea
and thus we are left with only three free parameters, namely
$m_{\overline{{\bf 5}}}\equiv m_{\overline{{\bf 5}}_3}$, 
$m_{{\bf 10}}\equiv m_{{\bf 10}_3}$
and $M$.

\subsection{\FUTB}
Also in the case of \FUTB\ the symmetry is enhanced after the reduction
of couplings.  The superpotential has now a 
  $Z_4\times Z_4\times Z_4$ symmetry with charges as shown in Table
\ref{tableB} and  with the
following superpotential
\bea
W &=& \sum_{i=1}^{3}\,[~\frac{1}{2}g_{i}^{u}
\,{\bf 10}_i{\bf 10}_i H_{i}+
g_{i}^{d}\,{\bf 10}_i \overline{\bf 5}_{i}\,
\overline{H}_{i}~] +
g_{23}^{u}\,{\bf 10}_2{\bf 10}_3 H_{4} \\
 & &+g_{23}^{d}\,{\bf 10}_2 \overline{\bf 5}_{3}\,
\overline{H}_{4}+
g_{32}^{d}\,{\bf 10}_3 \overline{\bf 5}_{2}\,
\overline{H}_{4}+
g_{2}^{f}\,H_{2}\, 
{\bf 24}\,\overline{H}_{2}+ g_{3}^{f}\,H_{3}\, 
{\bf 24}\,\overline{H}_{3}+
\frac{g^{\lambda}}{3}\,({\bf 24})^3~,\nonumber
\label{w-futb}
\eea
For this model the non-degenerate and isolated solutions to
$\gamma^{(1)}_{i}=0$ give us: 
\bea 
&& (g_{1}^{u})^2
=\frac{8}{5}~ g^2~, ~(g_{1}^{d})^2
=\frac{6}{5}~g^2~,~
(g_{2}^{u})^2=(g_{3}^{u})^2=\frac{4}{5}~g^2~,\label{zoup-SOL52}\\
&& (g_{2}^{d})^2 = (g_{3}^{d})^2=\frac{3}{5}~g^2~,~
(g_{23}^{u})^2 =\frac{4}{5}~g^2~,~
(g_{23}^{d})^2=(g_{32}^{d})^2=\frac{3}{5}~g^2~,
\nonumber\\
&& (g^{\lambda})^2 =\frac{15}{7}g^2~,~ (g_{2}^{f})^2
=(g_{3}^{f})^2=\frac{1}{2}~g^2~,~ (g_{1}^{f})^2=0~,~
(g_{4}^{f})^2=0~,\nonumber 
\eea 
and from the sum rule we obtain:
\bea
m^{2}_{H_u}+
2  m^{2}_{{\bf 10}} &=&M^2~,~
m^{2}_{H_d}-2m^{2}_{{\bf 10}}=-\frac{M^2}{3}~,~\nonumber\\
m^{2}_{\overline{{\bf 5}}}+
3m^{2}_{{\bf 10}}&=&\frac{4M^2}{3}~,
\eea
i.e., in this case we have only two free parameters  
$m_{{\bf 10}}\equiv m_{{\bf 10}_3}$  and $M$ for the dimensionful sector.

\begin{table}
\begin{center}
\renewcommand{\arraystretch}{1.3}
\begin{tabular}{|l|l|l|l|l|l|l|l|l|l|l|l|l|l|l|l|}
\hline
& $\overline{{\bf 5}}_{1} $ & $\overline{{\bf 5}}_{2} $& $\overline{{\bf
    5}}_{3}$ & ${\bf 10}_{1} $ &  ${\bf 10}_{2}$ &  ${\bf
  10}_{3} $ & $ H_{1} $ & $H_{2} $ & $ H_{3}
$ &$H_{4}$&   $\overline H_{1} $ & 
$\overline H_{2} $ & $\overline H_{3} $ &$\overline H_{4 }$&${\bf 24} $\\\hline
$Z_4$ & 1 & 0 & 0 & 1 & 0 & 0 & 2 & 0 & 0 & 0 & -2 & 0 & 0 & 0 &0  \\\hline
$Z_4$ & 0 & 1 & 0 & 0 & 1 & 0 & 0 & 2 & 0 & 3 & 0 & -2 & 0 & -3& 0  \\\hline
$Z_4$ & 0 & 0 & 1 & 0 & 0 & 1 & 0 & 0 & 2 & 3 & 0 & 0 & -2& -3 & 0 \\\hline
\end{tabular}
  \caption{Charges of the $Z_4\times Z_4\times Z_4$ symmetry for Model
    \FUTB.}
\label{tableB}
\renewcommand{\arraystretch}{1.0}
\end{center}

\end{table}

As already mentioned, after the $SU(5)$ gauge symmetry breaking we
assume we have the MSSM, i.e. only two Higgs doublets.  This can be
achieved by introducing appropriate mass terms that allow to perform a
rotation of the Higgs sector \cite{Leon:1985jm, finite1,Hamidi:1984gd,
  Jones:1984qd}, in such a way that only one pair of Higgs doublets,
coupled mostly to the third family, remains light and acquire vacuum
expectation values.  To avoid fast proton decay the usual fine tuning
to achieve doublet-triplet splitting is performed.  Notice that,
although similar, the mechanism is not identical to minimal $SU(5)$,
since we have an extended Higgs sector.

Thus, after the gauge symmetry of the GUT theory is broken we are left
with the MSSM, with the boundary conditions for the third family given
by the finiteness conditions, while the other two families are basically
decoupled.

We will now examine the phenomenology of such all-loop Finite Unified
theories with $SU(5)$ gauge group and, for the reasons expressed
above,   we will concentrate only on the
third generation of quarks and leptons. An extension to three
families, and the generation of quark mixing angles and masses in
Finite Unified Theories has been addressed in \cite{Babu:2002in},
where several examples are given. These extensions are not considered
here.  Realistic Finite Unified Theories based on product gauge
groups, where the finiteness implies three generations of matter, have
also been studied \cite{Ma:2004mi}.


\section{Restrictions from the low-energy observables}
\label{sec:ewpo}

Since the gauge symmetry is spontaneously broken below $M_{\rm GUT}$,
the finiteness conditions do not restrict the renormalization
properties at low energies, and all it remains are boundary conditions
on the gauge and Yukawa couplings (\ref{zoup-SOL5}) or
(\ref{zoup-SOL52}), the $h=-MC$ relation (\ref{hY}), and the soft
scalar-mass sum rule (\ref{zoup-sumr}) at $M_{\rm GUT}$, as applied in
the two models.  Thus we examine the evolution of these parameters
according to their RGEs up to two-loops for dimensionless parameters
and at one-loop for dimensionful ones with the relevant boundary
conditions.  Below $M_{\rm GUT}$ their evolution is assumed to be
governed by the MSSM.  We further assume a unique supersymmetry
breaking scale $M_{\rm SUSY}$ (which we define as the geometrical average
of the stop masses) and therefore below that scale the effective
theory is just the SM.  This allows to evaluate observables at or
below the electroweak scale.

In the following, we briefly describe the low-energy observables used in
our analysis. We discuss the current precision of
the experimental results and the theoretical predictions. 
We also give relevant details of the higher-order perturbative
corrections that we include. 
We do not discuss theoretical
uncertainties from the RG running between the high-scale parameters
and the weak scale.
At present, these uncertainties are expected to be 
less important than the experimental and theoretical uncertainties of
the precision observables. 

As precision observables we first discuss the 3rd generation quark
masses that are leading to the strongest constraints on the models under
investigation. Next we apply $B$~physics and Higgs-boson mass
constraints. Parameter points surviving these constraints are then
tested with the cold dark matter (CDM) abundance in the early
universe. We also briefly discuss the anomalous magnetic moment of the
muon.


\subsection{The quark masses}
\label{sec:mtmb}

Since the masses of the (third generation) quarks are no free
parameters in our model but predicted in terms of the GUT scale
parameters and the $\tau$~mass, $\mt$ and $\mb$ are (as it turns out the
most restrictive) precision observables for our analysis. For the
top-quark mass we use the current 
experimental value for the pole mass~\cite{mt1709}
\BE
\label{mtexp}
\mt^{\rm exp} = 170.9 \pm 1.8 \gev~.
\EE
For the bottom-quark mass we use the value at the bottom-quark mass
scale or at $\MZ$~\cite{pdg} 
\BE
\mbbar(\mb) = 4.25 \pm 0.1 \gev \mbox{~ ~or~ ~} 
\mbbar(\MZ) = 2.82 \pm 0.07 \gev~.
\EE
It should be noted that a numerically important correction appears in
the relation between the bottom-quark mass and the bottom Yukawa
coupling (that also enters the corresponding RGE running). The leading
$\tb$-enhanced corrections arise from one-loop contributions with
gluino-sbottom and chargino-stop loops. We include the leading effects
via the quantity $\De_b$~\cite{deltamb} (see also
\citeres{deltamb1,deltamb2,deltamb3}). Numerically the correction to the
relation between the bottom-quark mass and the bottom Yukawa coupling is
usually by far the dominant part of the contributions from the sbottom
sector (see also \citeres{mhiggsEP4,mhiggsFD2}). In the limit of large
soft SUSY-breaking parameters and $\tb \gg 1$, $\De_b$ is given
by~\cite{deltamb} 
\BE
\De_b = \frac{2\als}{3\,\pi} \, \mgl \, \mu \, \tb \,
                    \times \, I(\msbe, \msbz, \mgl) +
      \frac{\alt}{4\,\pi} \, \At \, \mu \, \tb \,
                    \times \, I(\mste, \mstz, |\mu|) ~,
\label{def:dmb}
\end{equation}
where the gluino mass is denoted by $\mgl$ and 
$\al_f \equiv h_f^2/(4\,\pi)$, $h_f$ being a fermion Yukawa coupling. 
The function $I$ is defined as
\BEA
I(a, b, c) &=& \ed{(a^2 - b^2)(b^2 - c^2)(a^2 - c^2)} \,
               \KL a^2 b^2 \log\frac{a^2}{b^2} +
                   b^2 c^2 \log\frac{b^2}{c^2} +
                   c^2 a^2 \log\frac{c^2}{a^2} \KR \\
 &\sim& \ed{\mbox{max}(a^2, b^2, c^2)} ~. \non
\EEA
A corresponding correction of \order{\al_{\tau}} has been included for
the relation between the $\tau$~lepton mass and the $\tau$ Yukawa
coupling. However, this correction is much smaller than the one given in
\refeq{def:dmb}. 

The $\De_b$ corrections are included by the replacement
\BE
\overline{m}_b \; \to \; \frac{\overline{m}_b}{1 + \De_b}~,
\label{dbresum}
\EE
resulting in a resummation of the leading terms in \order{\als\tb} and
\order{\alt\tb} to all-orders. Expanding \refeq{dbresum} to first or
second order gives an estimate of the effect of the resummation of the
$\De_b$ terms and has been used as an estimate of unknown higher-order
corrections (see below).


\subsection{The decay $b \to s \ga$}
\label{sec:bsg}

Since this decay occurs at the loop level in the SM, the MSSM 
contribution might {\it a priori} be of similar magnitude. A
recent
theoretical estimate of the SM contribution to the branching ratio at
the NNLO QCD level is~\cite{bsgtheonew}
\BE
\br( b \to s \ga ) = (3.15 \pm 0.23) \times 10^{-4}~.
\label{bsga}
\EE
It should be noted that the error estimate for $\br(b \to s \ga)$ is
still under discussion~\cite{hulupo}, and that other SM contributions to 
$b \to s \ga$ have been calculated~\cite{bsgneubert}. These
corrections are small compared with the theoretical uncertainty quoted
in \refeq{bsga}. 

For comparison, the present experimental 
value estimated by the Heavy Flavour Averaging Group (HFAG)
is~\cite{bsgexp,hfag}
\BE
\br(b \to s \ga) = (3.55 \pm 0.24 {}^{+0.09}_{-0.10} \pm 0.03) \times 10^{-4},
\label{bsgaexp}
\EE
where the first error is the combined statistical and uncorrelated systematic 
uncertainty, 
the latter two errors are correlated systematic theoretical uncertainties
and corrections respectively. 

Our numerical results have been derived with the 
$\br(b \to s \ga)$ evaluation provided in \citeres{bsgGH,ali,ali2},
incorporating also the latest SM corrections provided in~\citere{bsgtheonew}. 
The calculation has been checked against other
codes~\cite{bsgMicro,bsgKO1,bsgKO2}.
Concerning the total error in a conservative approach we add linearly
the errors of \refeqs{bsga} and (\ref{bsgaexp}) as well an intrinsic
SUSY error of $0.15 \times 10^{-4}$~\cite{ehoww}.


\subsection{The decay $B_s \to \mu^+\mu^-$}
\label{sec:bsmm}

The SM prediction for this branching ratio is $(3.4 \pm 0.5) \times
10^{-9}$~\cite{bsmmtheosm}, and 
the present experimental upper limit from the Fermilab Tevatron collider
is $5.8 \times 10^{-8}$ at the $95\%$ C.L.~\cite{bsmmexp}, still providing
the possibility for the MSSM to dominate the SM contribution. The
current Tevatron 
sensitivity, being based on an integrated luminosity of about 2~\ifb, 
is expected to improve somewhat in the future. In \citere{bsmmexp} an
estimate of the future Tevatron sensitivity of $2 \times 10^{-8}$ at
the 90\% C.L.\ has been given, and a sensitivity even down to the SM
value can be expected at the LHC. Assuming the SM value, i.e.\
$\br(B_s \to \mu^+ \mu^-) \approx 3.4 \times 10^{-9}$, it has been
estimated~\cite{lhcb} that LHCb can observe 33~signal events
over 10~background events within 3~years of low-luminosity
running. Therefore this process offers good prospects for probing the MSSM. 

For the theoretical prediction we use the code implemented in
\citere{bsgMicro} (see also \citere{bsmumu}), which includes the full
one-loop evaluation and the leading two-loop QCD corrections. We are
not aware of a detailed estimate of the theoretical uncertainties from
unknown higher-order corrections.


\subsection{The lightest MSSM Higgs boson mass}
\label{sec:mh}

The mass of the lightest $\cp$-even MSSM Higgs boson can be predicted in 
terms of
the other SUSY parameters. At the tree level, the two $\cp$-even Higgs 
boson masses are obtained as a function of $\MZ$, the $\cp$-odd Higgs
boson mass $\MA$, and $\tb$. 
We employ the Feynman-diagrammatic method for the theoretical prediction
of $\Mh$, using the code 
{\tt FeynHiggs}~\cite{feynhiggs,mhiggslong,mhiggsAEC,mhcMSSMlong}, which
includes all relevant higher-order corrections. 
The status of the incorporated results 
can be summarized as follows. For the
one-loop part, the complete result within the MSSM is 
known~\cite{mhiggsf1lB,mhiggsf1lC}. Concerning the two-loop
effects, their computation is quite advanced, see \citere{mhiggsAEC} and
references therein. They include the strong corrections
at \order{\al_t\als} and Yukawa corrections at \order{\al_t^2}
to the dominant one-loop \order{\al_t} term, and the strong
corrections from the bottom/sbottom sector at \order{\al_b\als}. 
For the $b/\Sbot$~sector
corrections also an all-order resummation of the $\Tb$-enhanced terms,
\order{\al_b(\als\tb)^n}, is known.
The current intrinsic error of $\Mh$ due to unknown higher-order
corrections have
been estimated to be~\cite{mhiggsAEC,mhiggsFDalbals,mhiggsWN,PomssmRep}
\BE
\De\Mh^{\rm intr,current} = 3 \gev . 
\label{eq:Mhintr}
\EE

The lightest MSSM Higgs boson is the models under consideration is
always SM-like (see also \citeres{asbs1,ehow1}). Consequently, the
current LEP bound of 
$\Mh^{\rm exp} > 114.4 \gev$ at the 95\%~C.L.\ can be taken
over~\cite{LEPHiggsSM,LEPHiggsMSSM}.


\subsection{Cold dark matter density}
\label{sec:cdm}

Finally we discuss the impact of the cold dark matter (CDM) density.
It is well known that the lightest neutralino, being the lightest
supersymmetric particle (LSP), is an 
excellent candidate for CDM~\cite{EHNOS}.
Consequently we demand that the lightest neutralino is indeed the
LSP.  Parameters leading to a different LSP are discarded.

The current bound, favored by a joint analysis of WMAP and other
astrophysical and cosmological data~\cite{WMAP}, is at the
$2\,\si$~level given by the range 
\BE
0.094 < \Omega_{\rm CDM} h^2 < 0.129~.
\label{cdmexp}
\EE
Assuming that the cold dark matter is 
composed predominantly of LSPs, the determination of
$\Omega_{\rm CDM} h^2$ imposes very strong constraints on the 
MSSM parameter space. 
As will become clear below, no model points fulfill the strict bound of
\refeq{cdmexp}. On the other hand, many model parameters would yield a very
large value of $\Omega_{\rm CDM}$. 
It should be kept in mind that somewhat larger values might be allowed
due to possible uncertainties in the determination of the SUSY spectrum
(as they might arise at large $\tb$, see below).

However, on a more general basis and not speculating about unknown
higher-order uncertainties, a mechanism is needed in our model to
reduce the CDM abundance in the early universe.  This issue could, for
instance, be related to another problem, that of neutrino masses.
This type of masses cannot be generated naturally within the class of
finite unified theories that we are considering in this paper,
although a non-zero value for neutrino masses has clearly been
established~\cite{pdg}.  However, the class of FUTs discussed here
can, in principle, be easily extended by introducing bilinear R-parity
violating terms that preserve finiteness and introduce the desired
neutrino masses \cite{Valle:1998bs}.  R-parity violation~\cite{herbi}
would have a small impact on the collider phenomenology presented here
(apart from fact the SUSY search strategies could not rely on a
`missing energy' signature), but remove the CDM bound of
\refeq{cdmexp} completely.  The details of such a possibility in the
present framework attempting to provide the models with realistic
neutrino masses will be discussed elsewhere.  Other mechanisms, not
involving R-parity violation (and keeping the `missing energy'
signature), that could be invoked if the amount of CDM appears to be
too large, concern the cosmology of the early universe.  For instance,
``thermal inflation''~\cite{thermalinf} or ``late time entropy
injection''~\cite{latetimeentropy} could bring the CDM density into
agreement with the WMAP measurements.  This kind of modifications of
the physics scenario neither concerns the theory basis nor the
collider phenomenology, but could have a strong impact on the CDM
derived bounds.

Therefore, in order to get an impression of the
{\em possible} impact of the CDM abundance on the collider phenomenology
in our models under investigation, we will analyze the case that the LSP
does contribute to the CDM density, and apply a more loose bound of  
\BE
\Omega_{\rm CDM} h^2 < 0.3~.
\label{cdmloose}
\EE
(Lower values than the ones permitted by \refeq{cdmexp} are naturally
allowed if another particle than the lightest neutralino constitutes
CDM.)
For our evaluation we have used the code {\tt MicroMegas}~\cite{bsgMicro}.


\subsection{The anomalous magnetic moment of the muon}
\label{sec:g-2}

We finally comment on the status and the impact of the anomalous
magnetic moment of the muon, $a_\mu \equiv \frac{1}{2} (g-2)_\mu$.
The SM prediction for $a_\mu$ 
(see~\citeres{g-2review,g-2reviewDS,g-2reviewFJ,LBLrev} for reviews) 
depends on the evaluation of QED contributions, the
hadronic vacuum polarization and light-by-light (LBL) contributions. 
The evaluations of the 
hadronic vacuum polarization contributions using $e^+ e^-$ and $\tau$ 
decay data give somewhat different results. 
The latest estimate based on $e^+e^-$ data~\cite{DDDD} is given by:
\BE
\amutheo = 
(11\, 659\, 180.5 \pm 4.4_{\rm had} \pm 3.5_{\rm LBL} \pm 0.2_{\rm QED+EW})
 \times 10^{-10},
\label{eq:amutheo}
\EE
where the source of each error is labeled. We note that the new $e^+e^-$
data sets that have recently been published in~\citeres{KLOE,CMD2,SND} have
been partially included in the updated estimate of $(g - 2)_\mu$. 

The SM prediction is to be compared with
the final result of the Brookhaven $(g-2)_\mu$ experiment 
E821~\cite{g-2exp2}, namely:
\BE
\amuexp = (11\, 659\, 208.0 \pm 6.3) \times 10^{-10},
\label{eq:amuexp}
\EE
leading to an estimated discrepancy~\cite{DDDD,g-2SEtalk}
\BE
\amuexp-\amutheo = (27.5 \pm 8.4) \times 10^{-10},
\label{delamu}
\EE
equivalent to a 3.3-$\sigma$ effect (see also
\citeres{g-2reviewFJ,g-2HMNT2,g-2reviewMRR}). 
In order to illustrate the possible size of
corrections, a simplified formula can be used, in which relevant
supersymmetric mass scales are set to a common value,
$\msusy = m_{\cha{}} = m_{\neu{}} = m_{\Smu} = m_{\Sneum}$. The result
in this approximation is given by
\BE
\amu^{\SU,{\rm 1L}} = 13 \times 10^{-10}
             \KL \frac{100 \gev}{\msusy} \KR^2 \tb\;
 {\rm  sign}(\mu).
\label{susy1loop}
\EE
It becomes obvious that $\mu < 0$ is already challenged by the
present data on $\amu$.
However, a heavy SUSY spectrum with $\mu < 0$ results in a $\amu^{\rm SUSY}$
prediction very close to the SM result. Since the SM is not regarded as
excluded by $(g-2)_\mu$, we also still allow both signs of $\mu$ in our
analysis. 

Concerning the MSSM contribution, the complete one-loop
result was evaluated a decade ago~\cite{g-2MSSMf1l}. 
In addition to the full one-loop contributions, the leading QED
two-loop corrections have also been
evaluated~\cite{g-2MSSMlog2l}. Further corrections at the two-loop
level have been obtained in \citeres{g-2FSf,g-2CNH}, 
leading to corrections to the one-loop result that are $\lsim 10\%$. These
corrections are taken into account in our analysis according to the
approximate formulae given in~\citeres{g-2FSf,g-2CNH}.


\section{Final Predictions}
\label{sec:finalpred}

In this section we present the predictions of the models \FUTA\ and \FUTB\
with ($\mu > 0$ and $\mu < 0$), whose theoretically restricted parameter
space due to finiteness has been further reduced by requiring 
correct electroweak symmetry breaking and the absence of charge or color
breaking minima. We furthermore demand that the 
bounds discussed in the previous section are also fulfilled, see the
following subsections.  We have
performed a scan over the GUT scale parameters, where we take as
further input the $\tau$ mass, $m_\tau = 1.777 \gev$.  This allows us to
extract the value of $v_u$, and then, using the relation 
$M_Z^2 = \frac{1}{2}\sqrt{(3g_1^2/5 + g_2^2)(v^2_u + v_d^2)}$, 
$v_{u,d} = 1/ \sqrt2 \langle H_{u,d} \rangle$, 
we can extract the value of $v_d$. In this way it is possible to predict
the masses of the top and bottom quarks, and the value of $\tb$.  As
already mentioned, we take into account the large radiative
corrections to the bottom mass, see \refeq{dbresum}, 
as well as the ones to the tau mass.
We have furthermore estimated the corrections to the top mass in our
case and found them to be negligible, so they are not included in our
analysis.  As a general result for both models and both signs of $\mu$
we have a heavy SUSY mass spectrum, and 
$\tb$ always has a large value of $\tb \sim 44-56$.


\subsection{Results vs.\ quark masses}

The first low-energy constraint applied are the top- and bottom-quark
masses as given in \refse{sec:mtmb}. 
In \reffi{fig:mbvsM} we present the predictions of the models
concerning the bottom quark mass.  The steps in the values for \FUTA\
are due to the fact that fixed values of $M$ were taken, while the
other parameters $m_5$ and $m_{10}$ were varied. However, this selected
sampling of the parameter space is sufficient for us to draw our
conclusions, see below.

We present the predictions for
$\mbbar(M_Z)$, to avoid unnecessary errors coming from the running
from $\MZ$ to the $\mb$ pole mass, which are not related to the
predictions of the present models.  As already mentioned in section
\ref{sec:mtmb}, we estimated the effect of the unknown higher order
corrections.  For such large values of $\tb$, see above, in the case
of \FUTB\ for the bottom mass they are $\sim 8\%$, whereas for
\FUTA\ they can go to $\sim 30\%$ (these uncertainties are slightly
larger for $\mu > 0$ than for $\mu < 0$).  Although these theoretical
uncertainties are not shown in the graphs, they have been taken into
the account in the analysis of $\mbbar$, by selecting only the values
that comply with the value of the bottom mass within this theoretical error.

From the bounds on the $\mbbar(\MZ)$ mass, we can see from
\reffi{fig:mbvsM} that the region $\mu > 0$ is excluded both for
\FUTA\ and \FUTB\, while for $\mu < 0$ both models lie partially within the
experimental limits.

In \reffi{fig:Mtvsm5} we present the predictions of the models \FUTA\
and \FUTB\ concerning the top quark pole mass.  We recall that the
theoretical predictions of $\Mt$ have an uncertainty of $\sim 4 \%$
\cite{Kubo:1995cg}. 
The current experimental value is given in \refeq{mtexp}.
This clearly favors \FUTB\, while \FUTA\ corresponds to $\Mt$ values that
are somewhat outside the experimental range, even taking theoretical
uncertainties into account. Thus $\Mt$ and $\mbbar(\MZ)$ together single
out \FUTB\ with $\mu < 0$ as the most favorable solution.
From \refse{sec:g-2} it is obvious that $\mu < 0$ is already
challenged by the 
present data on $\amu$.
However, a heavy SUSY spectrum as we have here (see above and
\refse{sec:spectrum}) with $\mu < 0$ results in a $\amu^{\rm SUSY}$
prediction very close to the SM result. Since the SM is not regarded as
excluded by $(g-2)_\mu$, we continue with our analysis
of \FUTB\ with $\mu < 0$.

\begin{figure}[htb!]
\vspace{1cm}
\centerline{\includegraphics[width=8cm,angle=0]{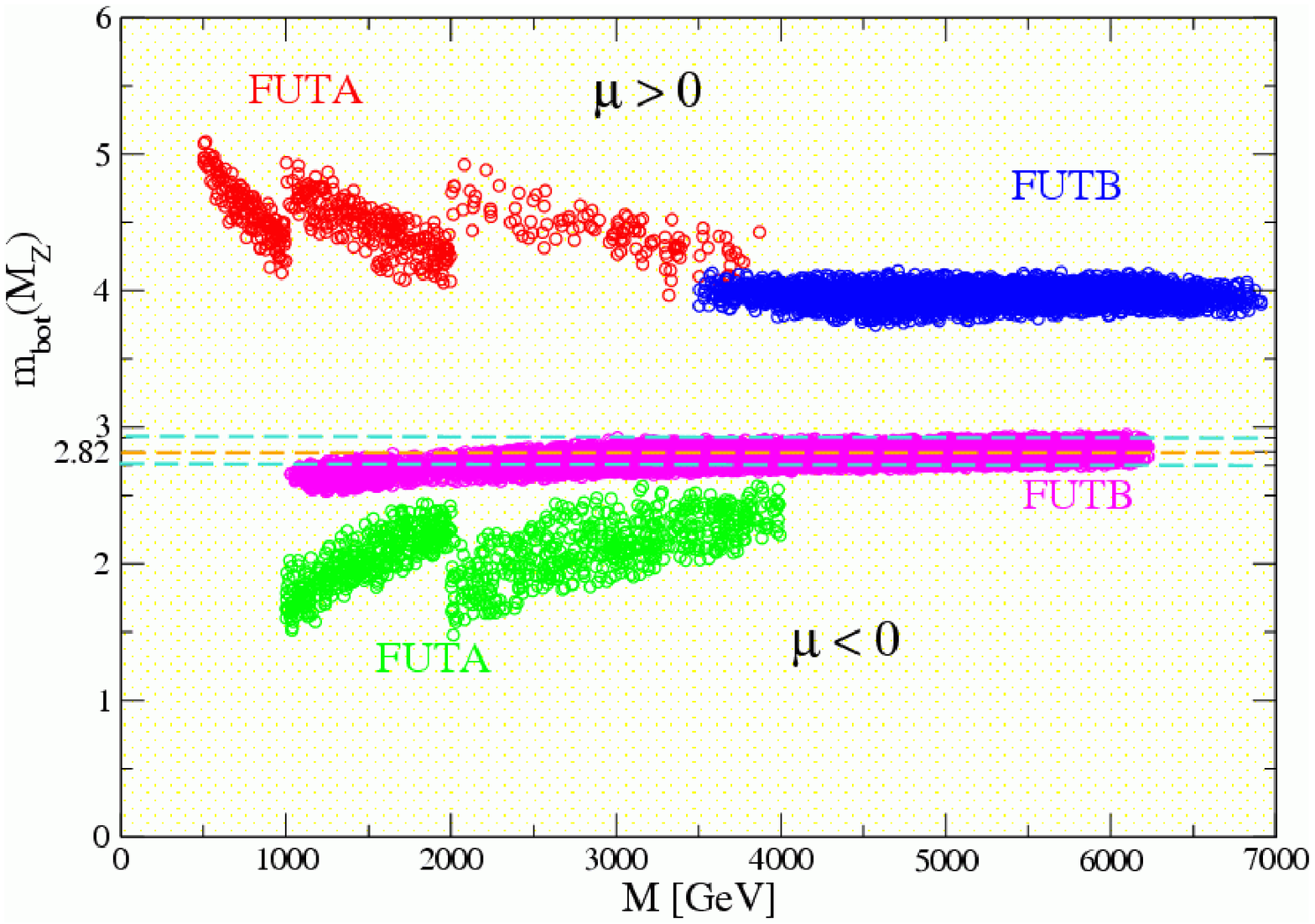}}
\caption{$\mbbar(\MZ)$ as function of $M$ for models $\bf FUTA$ and
$\bf FUTB$, for $\mu<0 $ and $\mu >0$.}
\label{fig:mbvsM}
\end{figure}


\begin{figure}[htb!]
\vspace{1cm}
\centerline{\includegraphics[width=8cm,angle=0]{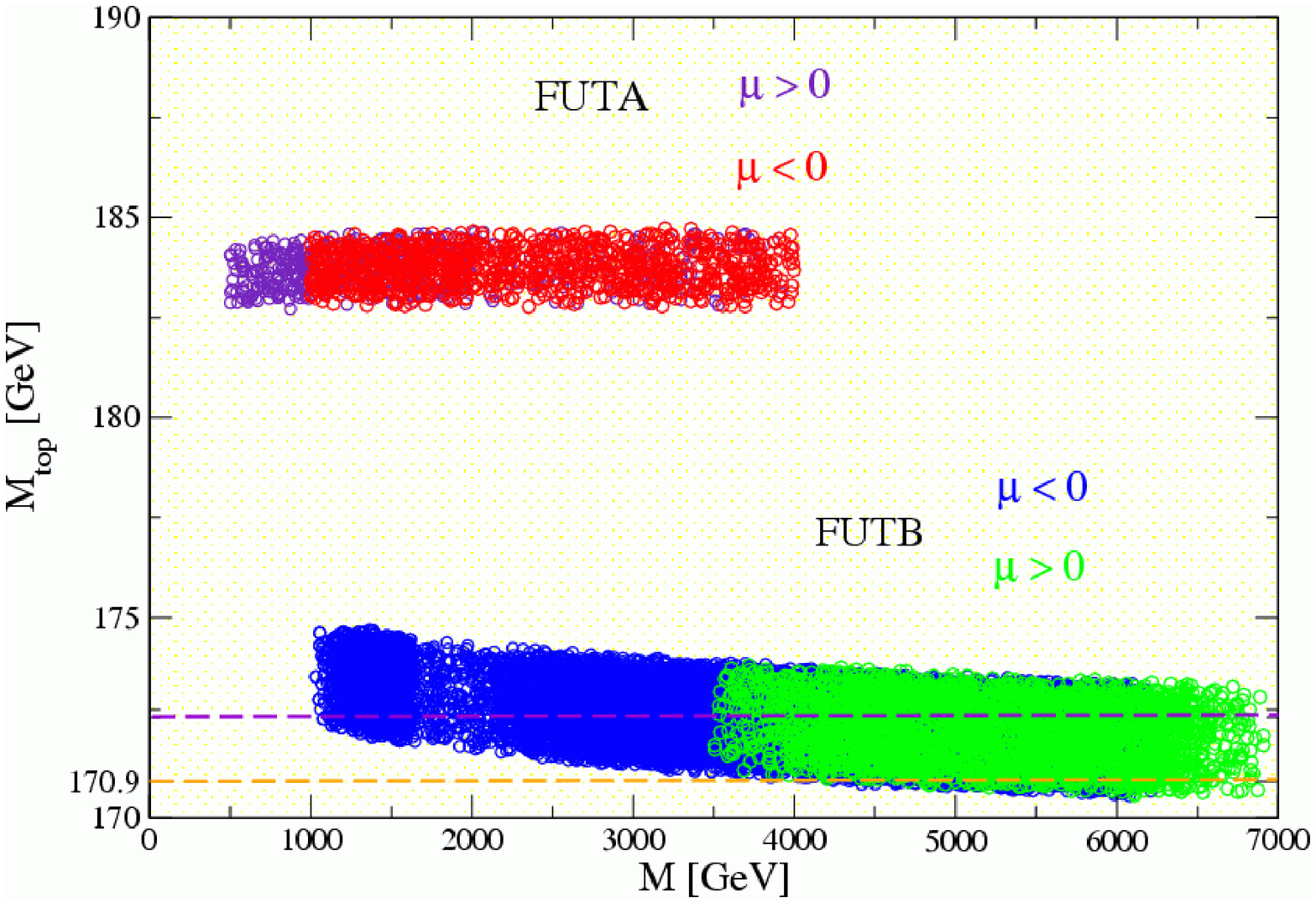}}
\caption{$\mt$  as function of $M$ for models $\bf FUTA$ and
$\bf FUTB$, for $\mu<0 $ and $\mu >0$.}
\label{fig:Mtvsm5}
\end{figure}


\subsection{Results for precision observables and CDM}

For the remaining model, \FUTB\ with $\mu < 0$, we compare the
predictions for $\br(b \to s \ga)$, $\br(B_s \to \mu^+\mu^-)$ and $\Mh$
with their respective experimental constraints, see \refses{sec:bsg}
-- \ref{sec:mh}.  
First, in \reffi{fig:BSGvsBMM} we show the predictions for 
$\br(b \to s \ga)$ vs.\ $\br(B_s \to \mu^+\mu^-)$ for all the points of
\FUTB\ with $\mu < 0$. The gray (red) points in the lower left corner
fulfill the $B$~physics constraints as given in \refses{sec:bsg},
\ref{sec:bsmm}. Shown also in black are the parameter points that
fulfill the loose CDM constraint of \refeq{cdmloose}, which can be found
in the whole $B$~physics allowed area.

\begin{figure}[htb!]
\vspace{10mm}
\centerline{\includegraphics[width=8cm,angle=0]{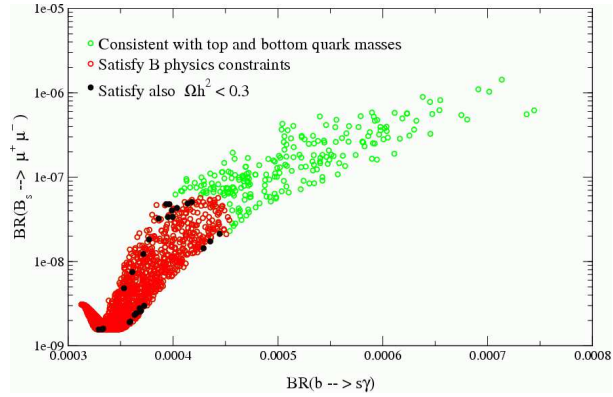}}
\caption{$\br(b \to s \ga)$ vs.\ $\br(B_s \to \mu^+\mu^-)$.  In green
  (light gray) are the points consistent with the top and bottom quark
  masses, in red (gray) are the subset of these
that fulfill the $B$~physics constraints, and in black the ones
that also satisfy the CDM loose constraint. }
\label{fig:BSGvsBMM}
\end{figure}

In the second step we test the compatibility with the Higgs boson mass
constraints and the CDM bounds. In \reffi{fig:higgsvsm5} we show $\Mh$
(as evaluated with 
{\tt FeynHiggs}~\cite{feynhiggs,mhiggslong,mhiggsAEC,mhcMSSMlong}) as a
function of $M$ for \FUTB\ with $\mu < 0$. Only the points that also
fulfill the $B$~physics bounds are included. The prediction for the Higgs
boson mass is constrained to the interval $\Mh = 118 \ldots 129 \gev$
(including the intrinsic uncertainties of \refeq{eq:Mhintr}),
thus fulfilling automatically the LEP bounds~\cite{LEPHiggsSM,LEPHiggsMSSM}.
Furthermore indicated in \reffi{fig:higgsvsm5} by the darker (red)
points is the parameter space that in addition
fulfills the CDM constraint as given 
in \refeq{cdmloose}. The loose bound permits values
of $M$ from $\sim 1000 \gev$ to about $\sim 3000 \gev$. 
The strong CDM bound, \refeq{cdmexp}, 
on the other hand, is not fulfilled by any data point, where the points
with lowest $\Omega_{\rm CDM} h^2 \sim 0.2$ can be found for 
$M \gsim 1500 \gev$. 
As mentioned in \refse{sec:cdm}, the CDM bounds should be viewed as
``additional'' constraints (when investigating the collider phenomenology). 
But even taking \refeq{cdmloose} at face
value, due to possible larger uncertainties
in the calculation of the SUSY spectrum as outlined above, the CDM
constraint (while strongly reducing the allowed parameter space) does not
exclude the model. Within the current calculation data points which are 
in strict agreement with \refeq{cdmexp} violate the $B$~physics constraints.

\begin{figure}[htb!]
\vspace{10mm}
\centerline{\includegraphics[width=8cm,angle=0]{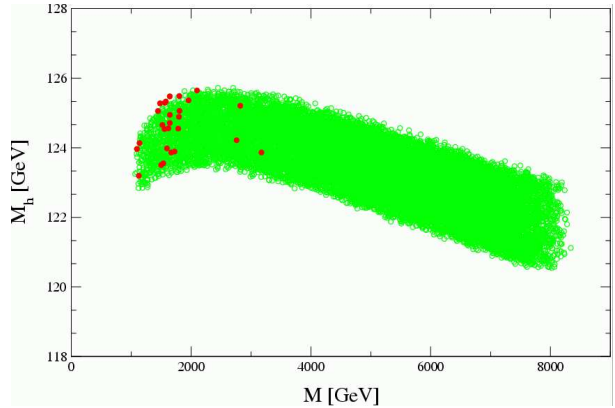}}
\caption{$\Mh$ is shown as a function of $M$. The light (green) points
  fulfill the $B$~physics constraints. The darker (red) dots in
  addition satisfy the loose CDM constraint of \refeq{cdmloose}.
}
\label{fig:higgsvsm5}
\end{figure}


\subsection{The heavy Higgs and SUSY spectrum}
\label{sec:spectrum}

The gray (red) points shown in \reffi{fig:BSGvsBMM} are the
prediction of 
the finite theories once confronted with low-energy experimental data.
In order to assess the discovery potential of the LHC~\cite{atlas,cms}
and/or the ILC~\cite{teslatdr,orangebook,acfarep,Snowmass05Higgs}
we show the corresponding predictions for the most relevant SUSY mass
parameters.  In \reffi{fig:LOSPvsm5} we plot the mass of the lightest
observable SUSY particle (LOSP) 
as function of $M$, that comply with the $B$~physics constraints, as
explained above.  
The darker (red) points fulfill in addition the loose CDM constraint
\refeq{cdmloose}.
The LOSP is either the light scalar $\tau$ or the second lightest
neutralino (which is close in mass with the lightest chargino). 
One can see that the masses are outside the reach of the LHC
and also the ILC. Neglecting the CDM constraint, even higher particle
masses are allowed.

\begin{figure}[htb!]
\vspace{10mm}
\centerline{\includegraphics[width=8cm,angle=0]{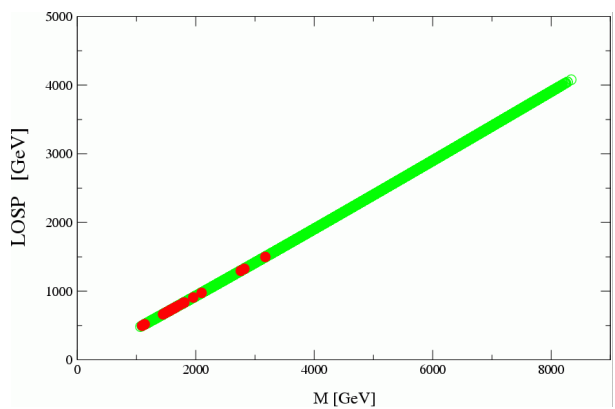}}
\caption{The mass of the LOSP is presented as a function of $M$.
Shown are only points that fulfill the $B$~physics constraints. The dark
(red) dots in addition also satisfy the loose CDM constraint of
\refeq{cdmloose}. 
}
\label{fig:LOSPvsm5}
\end{figure}

More relevant for the LHC are the colored particles. Therefore,
in \reffi{fig:LCSPvsm5} we show the masses of various colored 
particles: $\mste$, $\msbe$ and $\mgl$. The masses show
a nearly linear dependence on $M$. Assuming a discovery reach of 
$\sim 2.5 \tev$ yields a coverage up to $M \lsim 2 \tev$. This
corresponds to the largest part of the CDM favored parameter
space. 
All these particles are outside the reach of the ILC.
Disregarding the CDM bounds, see \refse{sec:cdm}, on the other hand,
results in large parts of the parameter space in which no SUSY
particle can be observed neither at the LHC nor at the ILC.

\begin{figure}[htb!]
\vspace{10mm}
\centerline{\includegraphics[width=8cm,angle=0]{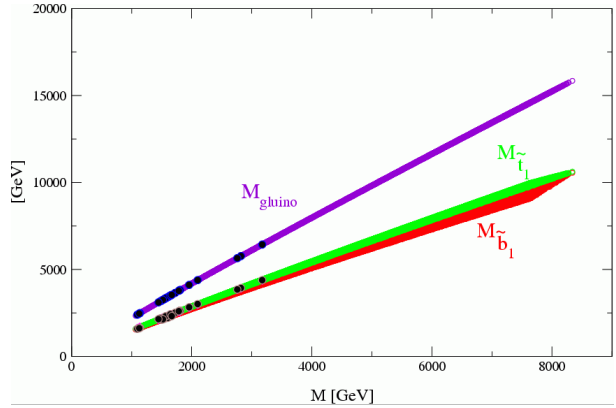}}
\caption{The mass of various colored particles are presented as a
  function of $M$. 
Shown are only points that fulfill the $B$~physics constraints, the
black  ones  satisfy also the loose CDM constraint. 
}
\label{fig:LCSPvsm5}
\end{figure}

We now turn to the predictions for the Higgs boson sector of \FUTB\ with
$\mu < 0$. 
In \reffi{fig:MAvsMh} we present the prediction for $\Mh$ vs.\ $\MA$, 
with the same color code as in \reffi{fig:LOSPvsm5}.
We have truncated the plot at about $\MA = 10 \tev$. The parameter
space allowed by $B$~physics extends up to $\sim 30 \tev$. 
The values that comply with the CDM constraints are in a relatively
light region of $\MA$ with $\MA \lsim 4000 \gev$. 
However, taking \reffis{fig:higgsvsm5} and \ref{fig:MAvsMh} into
account, the LHC and the ILC will observe only a light Higgs boson,
whereas the heavy Higgs bosons remain outside the LHC or ILC reach.

\begin{figure}[htb!]
\vspace{10mm}
\centerline{\includegraphics[width=8cm,angle=0]{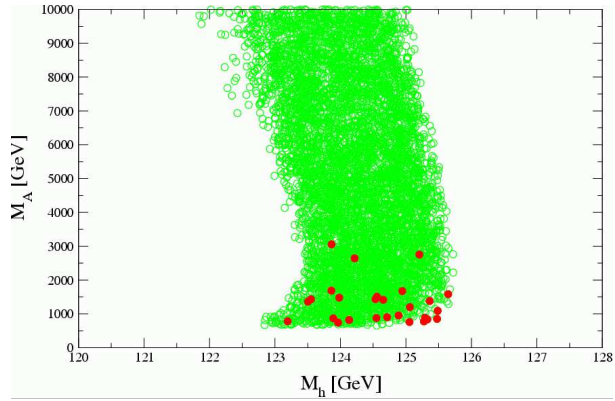}}
\caption{$M_A$ vs $\Mh$, with the same color code as in
  \reffi{fig:LOSPvsm5}.
}
\label{fig:MAvsMh}
\end{figure}

There might be the possibility to distinguish the light MSSM Higgs boson
from the SM Higgs boson by its decay characteristics. It has been shown
that the ratio 
\BE
\frac{\br(h \to b \bar b)}{\br(h \to WW^*)}
\EE
is the most powerful discriminator between the SM and the
MSSM using ILC measurements~\cite{deschi,ehhow}. We assume an
experimental resolution of this ratio of $\sim 1.5\%$ at the
ILC~\cite{barklow}.  
In \reffi{fig:BRvsM} we show the ratio as a function of $M$ with the
same color code as in \reffi{fig:LOSPvsm5}. It can be seen that up
to $M \lsim 2 \tev$ a deviation from the SM ratio of more than
$3\,\si$ can be observed. This covers most of the CDM favored
parameter space. Neglecting the CDM constraint, i.e.\ going to higher
values of $M$, results in a light Higgs boson that is
indistinguishable from a SM Higgs boson.

\begin{figure}[htb!]
\vspace{10mm}
\centerline{\includegraphics[width=8cm,angle=0]{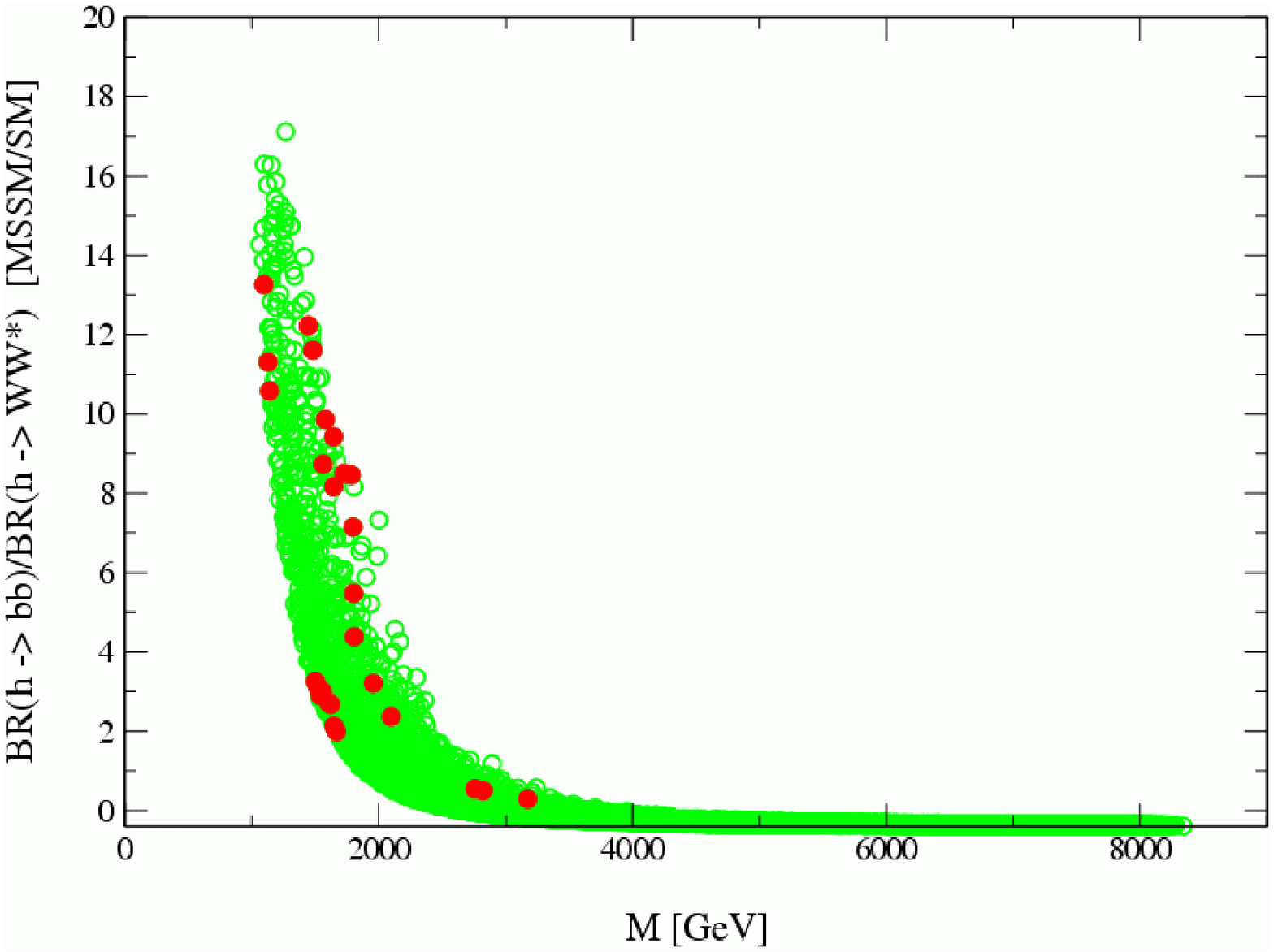}}
\caption{$\br(h \to b \bar b)/\br(h \to WW^*)$ [MSSM/SM] (expressed in
  terms of $\si$ with a resolution of $1.5\%$ (see text)) is shown 
as a function of $M$. The color code is the same as in 
  \reffi{fig:LOSPvsm5}.
}
\label{fig:BRvsM}
\end{figure}

Finally, in \refta{table1} we present a representative example of the
values obtained for the SUSY and Higgs boson masses for Model \FUTB\
with $\mu < 0$. The masses are typically large, as already mentioned,
with the LOSP starting from $\gsim 1000 \gev$.

 \begin{table}
\renewcommand{\arraystretch}{1.3}
\begin{center}
\begin{tabular}{|c|c||c|c|}
 \hline
$\mt  $ & 172  & $\mbbar(M_Z)$ & 2.7  \\ \hline
$\tb = $ & 46 & $\alpha_s $ & 0.116 \\ \hline\hline
$\mneu{1}$ & 796 & $m_{\tilde{\tau}_2}$ & 1268 \\ \hline
$\mneu{2}$ & 1462 & $m_{\tilde{\nu}_3}$ & 1575 \\ \hline
$\mneu{3}$ & 2048 & $\mu$  & -2046 \\ \hline
$\mneu{4}$ & 2052 & $B$  & 4722 \\ \hline
$\mcha{1}$ & 1462 & $\MA$ & 870 \\ \hline
$\mcha{2}$ & 2052  & $\MHp$ &  875 \\ \hline
$\mste$ & 2478 & $\MH$ & 869 \\ \hline
$\mstz$ & 2804 & $\Mh$ & 124 \\ \hline
$\msbe$ & 2513 & $M_1$ & 796 \\ \hline
$\msbz$ & 2783 & $M_2$ & 1467  \\ \hline
$m_{\tilde{\tau}_1}$ & 798 & $ M_3$ & 3655 
\\  \hline 
\end{tabular}
\end{center}
\renewcommand{\arraystretch}{1.0}
\caption{A representative spectrum of \FUTB\ with $\mu < 0$. All masses
  are in GeV.}
\label{table1}
\end{table}

\bigskip
It should be kept in mind that although we present the results that
are consistent with the (loose) CDM constraints, the present
model considers only the third generation of (s)quarks and
(s)leptons. A more complete analysis will be given elsewhere when
flavor mixing will be taken into account, see e.g.\
\citere{Babu:2002in}. A similar remark concerns the neutrino  
masses and mixings. It is well known that they can be introduced via 
bilinear R-parity violating terms \cite{herbi} which preserve finiteness.  
In this case the dark matter candidate will not be the lightest
neutralino, but could be another one, e.g.\ the axion.


\section{Conclusions}
\label{sec:conclusions}

In the present paper we have examined the predictions of two $N=1$
supersymmetric and moreover all-loop finite $SU(5)$ unified models,
leading after the spontaneous symmetry breaking at the Grand
Unification scale to the {\it finiteness-constrained MSSM}.

The finiteness conditions in the supersymmetric part of the unbroken
theory lead to relations among the dimensionless couplings, i.e.\ 
{\it gauge-Yukawa unification}. In addition the finiteness conditions
in the SUSY-breaking sector of the theories lead to a
tremendous reduction of the number of the independent soft SUSY-breaking
parameters leaving one model ({\bf A}) with three and another ({\bf B}) 
with two free parameters. 
Therefore the {\it finiteness-constrained MSSM}
consists of the well known MSSM with boundary conditions at the Grand
Unification scale for its various dimensionless and dimensionful
parameters inherited from the all-loop finiteness unbroken theories.
Obviously these lead to an extremely restricted and, consequently,
very predictive parameter space of the MSSM.  

In the present paper the
finiteness constrained parameter space of MSSM is confronted
with the existing low-energy phenomenology such as the top and bottom
quark masses, $B$~physics observables, the bound on the lightest Higgs
boson mass and constraints from the cold dark matter abundance in the
universe. In the first step the result of our parameter scan 
of the finiteness restricted parameter space of MSSM, after
applying the quark mass constraints and including theoretical
uncertainties at the unification scale, singles out the 
{\it finiteness-constrained MSSM} coming from the model ({\bf B}) with 
$\mu < 0$ (yielding $(g-2)_\mu$ values similar to the SM).
This model was further restricted by applying the $B$~physics
constraints. The remaining parameter space then automatically fulfills
the LEP bounds on the lightest MSSM Higgs boson with 
$\Mh = 118 \ldots 129 \gev$ (including already the intrinsic
uncertainties). In the final step the CDM measurements have been imposed. 
Considering the CDM constraints it should be kept in mind that
modifications in the model are possible (non-standard cosmology or
R-parity violating terms that preserve finiteness) that would have only
a small impact on the collider phenomenology. Therefore the CDM
relic abundance should be considered as an ``additional'' constraint,
indicating its {\em possible} impact.
In general, a relatively heavy SUSY and Higgs spectrum at the few~TeV
level has been obtained, where the lower range of masses yield better
agreement with the CDM constraint.
The mass of the lightest observable SUSY particle (the lightest
slepton or the second lightest neutralino) is larger than $500 \gev$,
which remains unobservable at the LHC and the ILC.
The charged SUSY particles start at around $1.5 \tev$ and grow nearly
linearly with $M$. Large parts of the CDM favored region results in
masses of stops and sbottoms below $\sim 2.5 \tev$ and thus 
might be detectable at the LHC.
The measurement of branching ratios of the lightest Higgs boson to
bottom quarks and $W$~bosons at the ILC shows a deviation to the SM
results of more than $3\,\si$ for values of $M \lsim 2.5 \tev$, again
covering most of the CDM favored region.

In conclusion, 
\FUTB\ with $\mu < 0$, fulfilling the existing constraints from 
quark masses, $B$~physics observables, Higgs boson searches and CDM
measurements, results at a heavy SUSY spectrum and large
$\tb$. Nonetheless, colored particles are likely to be observed in the
range of $\sim 2 \tev$ at the LHC. The ILC could measure a deviation
in the branching ratios of the lightest Higgs boson.
However, neglecting the CDM constraint allows larger values of
$M$. This results in a heavier SUSY spectrum, outside the reach of the
LHC and the ILC. In this case also the lightest Higgs boson is
SM-like.


\section*{Acknowledgements}

We acknowledge useful discussions with G.~Belanger, F.~Boudjema and
A.~Djouadi. We thank J.~Erler for discussions on $b$~quark mass
uncertainties. 
This work was supported by the EPEAEK programmes ``Pythagoras'' and
co-funded by the European Union (75\%) and the Hellenic state (25 \%); also
supported in part by the mexican grant PAPIIT-UNAM IN115207. 
Work supported in part by the European Community's Marie-Curie Research
Training Network under contract MRTN-CT-2006-035505
`Tools and Precision Calculations for Physics Discoveries at Colliders'.


\end{document}
